\documentclass[a4paper,USenglish]{lipics-v2016}

\usepackage{microtype}
\usepackage{url}
\usepackage{amssymb}
\usepackage{graphicx}
\usepackage{amsmath}
\usepackage{amsfonts}
\usepackage{amssymb}
\usepackage{amsthm}
\usepackage{color}

\title{Blockchain-based Proof of Location}

\author{Michele Amoretti, Giacomo Brambilla, Francesco Medioli, Francesco Zanichelli}
\affil{Department of Engineering and Architecture - University of Parma, Italy} 
\affil{Contact: michele.amoretti@unipr.it}
\authorrunning{M. Amoretti} 


\begin{document}

\maketitle

\begin{abstract}
Location-Based Services (LBSs) build upon geographic information to provide users with location-dependent functionalities. In such a context, it is particularly important that geographic locations claimed by users are trustworthy.
Centralized verification approaches proposed in the last few years are not satisfactory, as they entail a high risk to the privacy of users. In this paper, we present and evaluate a novel decentralized, infrastructure-independent proof-of-location scheme based on blockchain technology. Our scheme guarantees both location trustworthiness and user privacy preservation.
\end{abstract}

%%%%%%%%%%%%%%%%%%%%%%%%%%%%%%%%%%%%%%%%%%%
\section{Introduction}
\label{sec:introduction}

In recent years, an increasing number of Location-Based Services (LBSs) have been released, mostly because of the rapid expansion of the mobile device market \cite{Junglas2008}. 
LBSs take advantage of geographic location to provide users with accurate and targeted information for locating friends on a map, discovering nearby social events \cite{Wang2013}, generating alerts about traffic jams along a route \cite{Picone2012}, and more.

To ensure that such services work properly, it is necessary that geographic locations claimed by users are factual.
For example, LBSs with location-based access control that allow users to obtain a discount coupon, require that users cannot cheat on their position, to avoid delivering coupons to those who really have no right to obtain them.
Similarly, social networks that enable users to discover where their friends are, are meaningful only if geographic locations are trustworthy.

A \textit{proof of location} can be seen as a digital certificate that attests someone's presence at a certain geographic location, at a certain time. Different proof-of-location schemes have been proposed, either infrastructure-dependent or infrastructure-independent. It is worth nothing that most proof-of-location schemes are centralized, \textit{i.e.}, they rely on servers for storing proofs of location, which users have to trust either explicitly or implicitly.

With the objective of achieving a system that, at the same time, provides verification of geographic location of its users and ensures a high level of privacy to them, we have designed a completely decentralized and infrastructure-independent proof-of-location scheme for LBS-oriented peer-to-peer networks (like Overdrive \cite{Heep2013} or ADGT \cite{Brambilla2014-2}). 
The decentralized nature of peer-to-peer systems guarantees higher privacy levels, as it removes the central authority knowing both the geographic location of users and the information they exchange. In the following, peer-to-peer network, overlay network, peer-to-peer overlay and network are equivalent expressions we use with reference to the same concept.

Our proof-of-location scheme is based on \textit{blockchain} technology \cite{Nakamoto2008}. A blockchain is a cryptographically secure distributed ledger that maintains a continuously growing list of ordered data blocks. Each block header contains the hash of the previous block header, a hash value related to its data and a timestamp. Once recorded, the data in a block cannot be altered retroactively.

The main feature that differentiates the blockchain from all other distributed databases is its completely decentralized nature, which escapes the presence of a trusted central authority.
Indeed, blockchain maintenance is performed by a network of communicating nodes, which validate transactions, add them to their own local copy of the blockchain, and then broadcast block additions to other nodes. 
All these operations are performed in such a way that \textit{distributed consensus} emerges among network nodes, about the information stored in the blockchain.
If a forged block (\textit{i.e.}, a block with false proofs of location) is added to the chain, other nodes will find the data to be untrue. In the Bitcoin virtual currency system (www.bitcoin.org), the blockchain is used to store transaction records \cite{Nakamoto2008}. Ethereum (www.ethereum.org) is another virtual currency system with a public blockchain, which stores not only transactions, but also programs denoted as smart contracts.

Both Bitcoin and Ethereum use a proof of work (PoW) algorithm that rewards participants who solve cryptographic puzzles in order to validate transactions and create new blocks (\textit{i.e.}, mining). However, Ethereum is moving toward proof of stake (PoS), a category of consensus algorithms for public blockchains that depend on a validator's economic stake in the network. More precisely, a set of validators take turns proposing and voting on the next block, and the weight of each validator's vote depends on the size of its deposit (\textit{i.e.}, stake).

In the proposed scheme, a customized PoS-based blockchain is used to store proofs of location. The benefits of PoS as opposed to PoW are, in short: 
\begin{enumerate} 
\item no need to consume large quantities of electricity in order to secure a blockchain; 
\item enables a wider array of techniques that use game-theoretic mechanisms in order to better discourage Sybil groups from forming and, if they do form, from acting in ways that are harmful to the network; 
\item reduced centralization risks due to the uniform distribution of block validation workload; 
\item ability to use economic penalties against malicious players (as they are always exposed to the risk to loose their stake).
\end{enumerate}

The rest of the paper is organized as follows. In Section \ref{sec:related}, we describe related work. In Section \ref{sec:scheme}, we illustrate our proof-of-location scheme. In Section \ref{sec:analysis}, we analyze the proposed scheme in terms of robustness against various kinds of attacks. In Section \ref{sec:evaluation}, we illustrate a preliminary performance evaluation of the proposed scheme. Finally, we conclude the paper with a summary of achieved results and an outline for future work in Section \ref{sec:conclusion}.

%%%%%%%%%%%%%%%%%%%%%%%%%%%%%%%%%%%%%%%%%%%
\section{Related Work}
\label{sec:related}

The production of proofs of location with mobile devices may be either infrastructure-dependent or infrastructure-independent. 

In the infrastructure-dependent approach, a set of WiFi access points (APs) are assumed to be available to produce proofs of location to users.

Javali \textit{et al.} \cite{Javali2016} proposed a proof of location generation and verification scheme which proves the presence of the user within an area of interest at a particular time and the claim is securely verifiable by LBSs. The scheme is based on channel state information (CSI) obtained from 802.11 WiFi packet traces and on the fuzzy vault cryptographic primitive. A server stores the information required to verify the location claims submitted by the users.

Li \textit{et al.} \cite{Li2016} identified a new kind of attack coined as location cheating attack in database-driven cognitive radio networks (CRNs), which allows attackers to mislead honest users with a fake location and make them query the database with fake locations, or allows malicious users to claim a location arbitrarily and query the database for service. The authors proposed an infrastructure-based scheme that relies on the existing WiFi AP network or cellular network to provide guarantees for location cheating prevention and user location privacy. 

In the infrastructure-independent approach, users receive location claims from neighbors.

Liu \textit{et al.} \cite{Liu2015} proposed a scheme denoted as CLIP, which maintains user location privacy, allowing one to submit a portion of her/his mobility trace, with which the commitment to follow a specific path can be verified. Wireless APs or co-located mobile devices are used to generate proofs of location. CLIP uses a lightweight spatio-temporal trust model to detect false proofs of location deriving from collusion attacks. A prototype implementation on Android demonstrates that CLIP requires low computational and storage resources.

Another infrastructure-independent scheme is APPLAUS \cite{Zhu2013}, envisaging colocated Bluetooth-enabled users that mutually generate proofs of location and report them to a server. Although this solution appears effective and robust, its centralized architecture eases tracking of pseudonym-identified users by malicious administrators, whereas it might hinder the deployment of user-created LBSs.

%%%%%%%%%%%%%%%%%%%%%%%%%%%%%%%%%%%%%%%%%%%
\section{Proof-of-Location Scheme}
\label{sec:scheme}

In this section, we illustrate our proof-of-location scheme, by detailing the architecture of the system, the blockchain construction mechanism, and how distributed consensus is achieved.

\subsection{Architecture}

\begin{figure}[h!]
  \centering
  \includegraphics[width=7cm]{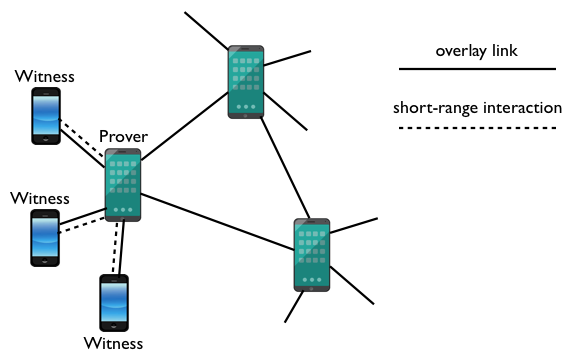}
  \caption{The envisioned LBS-oriented peer-to-peer network.}
  \label{fig:architecture}
\end{figure}

We consider an LBS-oriented peer-to-peer network with mobile nodes that are connected to the Internet through the WiFi or cellular network interface, and are able to exchange information with neighboring nodes through short-range communication technologies such as Bluetooth (Figure \ref{fig:architecture}), the de facto standard for personal networking, and 802.11p/ITS-G5 for vehicle-to-vehicle and vehicle-to-infrastructure interaction. 
A \textit{Prover} is a node that wants to collect proofs of location from its neighbors.
A \textit{Witness} for a Prover is a node that has provided a proof of location to the Prover. 
Every peer $i$ is described by a unique identifier $K^{pu}_{i}$, which is also its \textit{public key}.
Moreover, every peer $i$ is able to digitally sign messages with the \textit{private key} $K^{pr}_{i}$.

\subsection{Blockchain Construction}
\begin{figure}[h!]
  \centering
  \includegraphics[width=0.96\columnwidth]{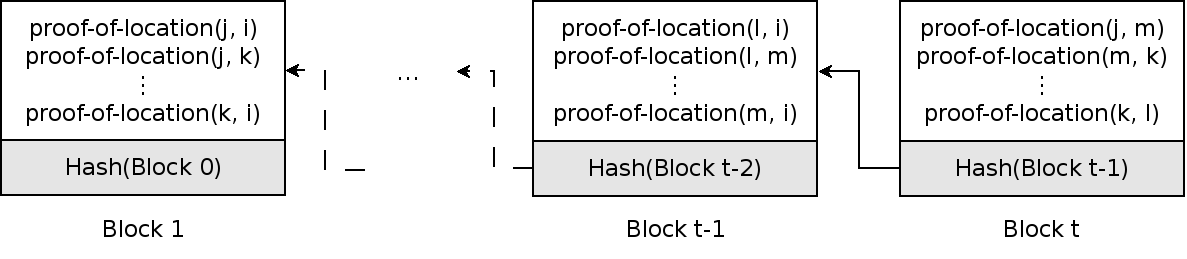}
  \caption{Proofs of location recorded in blocks of a blockchain. Every block contains a hash of the previous block.}
  \label{fig:blockchain}
\end{figure}

We have adopted a blockchain-based approach to endow networked nodes with the capability to verify and store proofs of location, not requiring a centralized supernode that oversees sensitive data of other nodes.
In our approach, recent valid proofs of location are recorded into blocks, which are then added to the end of the chain and, once confirmed by consensus, they cannot be changed, as shown in Figure \ref{fig:blockchain}.

\begin{figure}[h!]
  \centering
  \includegraphics[width=8cm]{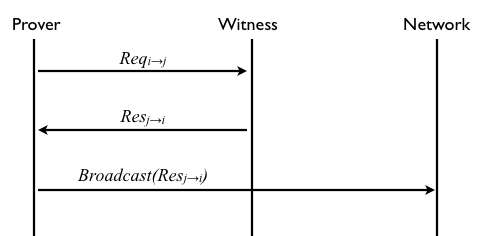}
  \caption{Construction and diffusion of a proof of location.}
  \label{fig:protocol}
\end{figure}

Similarly to the solution proposed by Zhu and Cao \cite{Zhu2013}, peers can communicate with near nodes through any effective short-range communication technology and they periodically use these interfaces to broadcast proof-of-location requests and responses to their neighbors, as illustrated in Figure \ref{fig:protocol}.

Supposing next block to be created is $Block_t$, a proof-of-location request from peer $i$ (the Prover) to peer $j$ (the Witness) contains the identifier $K^{pu}_{i}$ and the geographic location of peer $i$, as well as a hash of the latest created block $h(Block_{t-1})$.
As depicted in Figure \ref{fig:request}, the request is signed with the requester's private key, so that anyone can verify that it has not been tampered with.

\begin{figure}[ht]
  \begin{align*}
    Req_{i \rightarrow j}:
    \begin{Bmatrix}
      K^{pu}_{i} \\
      \langle latitude, longitude \rangle _{i} \\
      h(Block_{t-1}) \\
      timestamp
    \end{Bmatrix}_{K^{pr}_{i}}
  \end{align*}
  \caption{A proof-of-location request from peer $ i $ (the Prover) to peer $ j $ (the Witness).}
  \label{fig:request}
\end{figure}

The Witness (\textit{i.e.}, peer $j$) verifies the validity of the proof-of-location request, according to the following rules:
\begin{enumerate}
  \item the request has to come from a peer that, beyond being in touch through the short-range communication technology, is a known contact in the LBS-oriented peer-to-peer overlay network; 
  \item the request is produced and digitally signed by the Prover;
  \item the request contains an admissible geographic location, \textit{i.e.}, not farther than the adopted maximum distance reachable with the short-range communication technology;
  \item the request refers to the end block of the blockchain; otherwise, peers start a synchronization process via the peer-to-peer network to align their blockchains.
\end{enumerate}

Once all checks have been fulfilled, a proof-of-location response is produced by the Witness, wrapping the received request into a new message, together with its geographic location and identifier (\textit{i.e.}, its public key $K^{pu}_{j}$).
The proof-of-location response is also signed with the private key of the Witness, as illustrated in Figure \ref{fig:response}.

\begin{figure}[ht]
  \begin{align*}
    Res_{j \rightarrow i}:
    \begin{Bmatrix}
      Req_{i \rightarrow j} \\
      K^{pu}_{j} \\
      \langle latitude, longitude \rangle_{j} \\
      timestamp
    \end{Bmatrix}_{K^{pr}_{j}}
  \end{align*}
  \caption{A proof-of-location response produced and signed by peer $ j $ (the Witness) for peer $ i $ (the Prover).}
  \label{fig:response}
\end{figure}

Then, the response is verified by the Prover according to the following rules:
\begin{enumerate}
  \item the response comes from one of the peers to whom the request was sent;
  \item the response is digitally signed by the peer that produced it;
  \item the response contains an admissible geographic location, \textit{i.e.}, not farther than the maximum distance that is reachable with the adopted short-range communication technology.
\end{enumerate}

In case the response is correctly verified, it corresponds to a proof of location, attesting that two peers are geographically close to each other and specifying their geographic locations.
Proofs of location are then broadcasted to the network, which records them in the public ledger of all proofs of location, \textit{i.e.}, the blockchain, after validating them.

When a peer receives a proof of location declaring that another peer is located nearby, yet that peer is neither reachable by short-range communication nor belonging to the list of known neighbors, the proof of location is discarded and not further forwarded within the network.

Conversely, when a peer receives a proof of location declaring that another peer is located outside the area covered by the neighborhood monitoring protocol, the former peer may either immediately discard the proof of location (conservative approach) or use the blockchain to compute the \textit{betweenness} $B$ of the two peers that produced the proof of location in the \textit{pseudonym correlation graph}, over a limited but meaningful time period. 
The pseudonym correlation graph is an undirected graph whose vertices are the pseudonyms of the peers and each edge denotes that there is a proof of location between the corresponding vertex pair. Betweenness of vertex $v$ is defined as the number of shortest paths from all vertices to all others that pass through node $v$. On unweighted graphs like the pseudonym correlation one, calculating betweenness centrality takes $O(|V||E|)$ time using Brandes' algorithm \cite{Brandes2001}, where $|V|$ is the number of vertices and $|E|$ is the number of edges.  
As suggested by Zhu and Cao \cite{Zhu2013}, any peer with low $B$ may be considered as malicious and its proofs of location discarded. 

Every peer in the network puts all known valid unacknowledged proofs of location into a block, together with a reference to the previous valid block known to that peer.
In addition to proofs of location and the reference to its predecessor, the block contains the identifier of the peer that generated it.
Moreover, the block is signed with the private key of the peer that generated it, as shown in Figure \ref{fig:block}.

\begin{figure}[ht]
  \begin{align*}
    Block_{t}:
    \begin{Bmatrix}
      Res_{j \rightarrow i} \\
      Res_{j \rightarrow k} \\
      \vdots \\
      Res_{k \rightarrow i} \\
      K^{pu}_{i} \\
      h(Block_{t-1})
    \end{Bmatrix}_{K^{pr}_{i}}
  \end{align*}
  \caption{$t$-th block, produced and signed by the peer $i$.}
  \label{fig:block}
\end{figure}

Afterwards, the newly created block is broadcasted to the peers of the network, which decide whether to add the block to the end of the blockchain or not. If most peers add the block to the blockchain, then consensus is achieved, therefore proofs of location are made persistent.
Otherwise, the block is discarded and not attached to the blockchain.

Whereupon, it is verified that the hash of the referenced block matches the end block in the chain, otherwise a \textit{fork} in the blockchain occurs.
Which one branch will become part of the main blockchain depends on the distributed consensus algorithm explained below.

Last but not least, every peer makes sure that proofs of location specified in a new block are not already present in previous blocks of the blockchain. 
In case a proof of location concerns the geographic location of the peer itself, it is checked that signatory peers of the proof of location are known (\textit{i.e.}, they belong to the contact list provided by the peer-to-peer overlay).
If these conditions are not respected, the block is discarded, instead of being propagated into the network.

\subsection{Distributed Consensus}
\label{sec:consensus}

In the proposed system, the blockchain is built by means of a PoS approach, whereby, to decide the peer that must create next block in the blockchain, a pseudo-random selection is performed, taking into account the number $n_T$ of proofs of location in the latest $T$ blocks of the blockchain. The larger the $n_T$, the higher the probability for a peer to be chosen. 

No time-consuming and energy-hungry work is required for creating valid blocks. 
If a peer receives more than one valid block from its neighbors, it will add to the end of its blockchain the block produced by the peer with the largest $n_T$.
The latest $T$ blocks of the chain cannot include more than one block produced by the same peer. This is to prevent the monopoly problem, \textit{i.e.}, a peer that keeps out the proofs of location that concern other peers from the block it produces, in order to remain the owner of most proofs of location and, therefore, to take control of the blockchain.   

\begin{figure*}[t]
  \centering
  \includegraphics[width=0.96\textwidth]{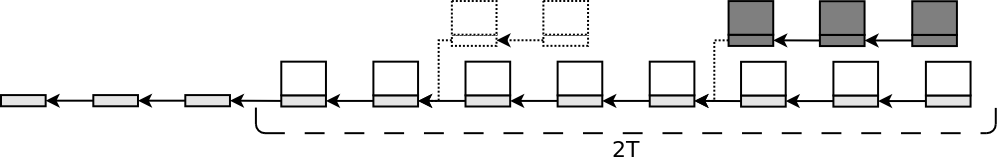}
  \caption{Valid proofs of location are persisted in the main blockchain (white blocks). Proofs of location that precede the latest $2T$ blocks are no longer significant. Grey blocks compose a fork competing to become the main blockchain. Dashed blocks are part of a past fork that has become invalid and ignored.}
  \label{fig:fork}
\end{figure*}

Preserving all block headers is sufficient to enable a simplified verification of the blockchain. Peers are not forced to maintain a copy of all proofs of location stored in the blockchain (similarly to lightweight Bitcoin clients \cite{Gervais2014}). The distributed consensus procedure requires that peers preserve the latest $T$ blocks. In order to handle forks of the blockchain, it is prudent to maintain the latest $2T$ blocks. The blocks that precede the latest $2T$ blocks may be \textit{pruned}, as depicted in Figure \ref{fig:fork}.  

The value of $T$ is application-dependant. When it is important to store several past geographic locations, such as in applications for tracking and monitor vehicle fleets, $T$ has a larger value, compared to applications that localize nearby friends.
Forensic applications may be interested to store the whole blockchain, in order to provide effective and trusted alibis for people under investigation.
On the other hand, the lower is the value of $T$, the smaller is also the space occupied in memory.
Apart from this, the protocol is independent from the application layer and highly versatile, supporting the realization of a wide range of LBSs.

%%%%%%%%%%%%%%%%%%%%%%%%%%%%%%%%%%%%%%%%%%%
\section{Robustness Analysis}
\label{sec:analysis}

In the following, we analyze the robustness of the proposed scheme with respect to all major LBS-related attacks \cite{Khan2014}. Once detected, malicious peers may be penalized by the community of honest peers. For example, an honest peer may generate a special-purpose proof-of-location attestation to inform other peers it will distrust future activities of the malicious peer.

\subsection{Cheating on own geographic location} 
A peer could declare a false geographic location, in order to obtain a false proof of location. Our scheme prevents this kind of attack, since each peer that receives a proof location request or response verifies that the specified geographic location is not farther than the maximum distance covered by the short-range communication technology.

A user with multiple identities (\textit{i.e.}, controlling multiple peers) may produce false proof of location in a selfish way. However, betweenness analysis, illustrated in Section \ref{sec:scheme}.B, will allow honest peers to detect such kind of malicious behavior.
  
\subsection{Cheating on another peer's geographic location} 
Another possible attack could hail from a peer that produces false claims about other peers' geographic locations. Our scheme precludes such an attack, thanks to the asymmetric cryptography mechanism, whereby all the declarations concerning geographic locations stated by peers are digitally signed with their private keys and easily verifiable using their public keys that correspond to their identifiers.

\subsection{Replaying proofs of location} 
Outdated proofs of location could be re-broadcasted in the network by malicious peers, with the purpose to forge the geographic location of other peers. Since every peer of the network checks that the proof of location is not already contained inside the blockchain before retransmitting it, it is not possible to successfully complete this attack. Moreover, inasmuch every proof of location contains a reference to a block of the blockchain, it is immediately discarded in case the referenced block is older that the latest $2T$ blocks of the blockchain.

\subsection{Colluding with other peers} 
Another threat exists when two or more malicious peers collude to generate false proofs of location. In literature, this kind of attack is denoted as \textit{Sybil attack} \cite{Douceur2002}. Let us consider a malicious peer that tries to prove itself in a geographic location that is not the actual one, with the help of another malicious peer. The two peers agree upon producing a proof of location attesting that their geographic locations are different from the actual ones. Then, they broadcast the false proof of location into the network.
  
In most cases, colluding peers can be detected by honest peers, thanks to the short-range communication technology. Moreover, it is unlikely that the whole list of peers provided by the peer-to-peer neighborhood monitoring protocol is made of colluding peers. For the sake of precision, the following three scenarios are possible.
\begin{enumerate}
    \item Proof of location and location declared in the peer-to-peer overlay are identical and both false. If a peer receives a proof of location concerning two other peers that claim to be close to it, it verifies that at least one of the two peers can be contacted with the short-range communication technology; if not, the proof of location is discarded. 
    \item Proof of location and location declared in the peer-to-peer overlay are different; one of them or both are false. If a peer receives from another peer a proof of location concerning the latter peer and related to a geographic location that is different from the one provided by the peer-to-peer overlay, such a proof of location is immediately discarded.
    \item Two peers collude to build a false proof of location for one of them. The proof may be received by a honest peer that is far from the considered location and cannot contact the colluding nodes with the short-range communication technology. In this case, the honest peer may either immediately discard the proof of location (conservative approach) or make a more contemplated decision, by evaluating the betweenness of the two suspect peers, according to the procedure illustrated in Subsection 3.2.
\end{enumerate}
Hence, collusion is hindered by information provided by peers belonging to the LBS-oriented peer-to-peer overlay.  

Moreover, large groups of colluding peers (Sybil groups) occupying a specific geographical space and discarding all proof of location requests from honest peers, would be be easily detected. Indeed, colluding peers should move in a coordinated way, which is not easy especially for very large groups. On the other hand, if they don't move, they would be highly suspect.

\subsection{Determining real identities of peers}
An attacker could attempt to determine the real identity of peers through full observation of proofs of location in the blockchain.
Actually, in our scheme there is no limit on the number of identifiers. In the same way as Bitcoin users are allowed to adopt different receiving addresses, our users can freely decide to change their peer identifiers. As proved by Zhu and Cao \cite{Zhu2013}, if a peer has the possibility to periodically change its identifier according to a Poisson distribution, it gains unobservability and an attacker cannot determine the real identity of the peer by observing location proof records.

\subsection{Dealing with highly dynamic environments}
Suppose to have two close peers located inside two different vehicles. When the proof-of-location request is sent by one of the two peers, the vehicles are in contact (in the short-range). But, when the reply is computed, the Witness is just outside the range of transmission. To avoid such kind of problem, the peer that issues the request should use the peer-to-peer overlay to detect the most suitable Witness among the moving peers that will be met with high probability, and send the proof-of-location request in suitable advance.

%%%%%%%%%%%%%%%%%%%%%%%%%%%%%%%%%%%%%%%%%%%
\section{Performance Evaluation}
\label{sec:evaluation}

Using OSMobility \cite{Brambilla2014-3}, a simulation platform for studying distributed/mobile systems within realistic geographical spaces, we have made a preliminary performance evaluation of our blockchain-based proof-of-location scheme.

The simulated scenario consists of a network with 100 peers, each one monitoring a circular area with a radius of 2 km by means of the ADGT protocol. The fraction of correctly pinpointed neighbors (Coverage Percentage) is $75$\%, in line with previous results by Brambilla \textit{et al.} \cite{Brambilla2014-2}. Moreover, the radius of the simulated short-range communication technology is $150$ m.
A fraction $P_1$ of peers act as Witnesses, \textit{i.e.}, produce and broadcast the proofs of location. 
A fraction $P_2$ of these Witnesses represent a group of colluding cheaters.

Cheating behavior is:
\begin{enumerate}
\item create a proof for a false location, with the help of a cheater that is located nearby that false location, and geocast the resulting proof of location;
\item manage received messages as follows:
\begin{itemize}
\item if the message is a denial of location, either discard it (full-cheating behavior) or propagate it (less-cheating behavior);
\item if the message is a proof of location produced by other cheaters, propagate it without checking it; 
\item if the message is a proof of location produced by a honest peer, verify that it is consistent with the blockchain; if it is, propagate it, otherwise turn it in a denial of location and propagate it.
\end{itemize}
\end{enumerate}

Conversely, upon receiving a proof of location declaring that another peer is located outside the area covered by the neighborhood monitoring protocol, honest peers immediately discard the proof of location (conservative approach).

In our simulations, we have considered $P_1 \in \{10,30,50,100\}$ [\%] and $P_2 \in \{0,25,50,75,100\}$ [\%].
Witnesses (including cheaters) have been randomly selected according to a uniform spatial distribution.

We have measured the following performance indicators:
\begin{enumerate}
  \item $TP$ [\%], the percentage of true proofs of location that are stored into the blockchain, with respect to the total number of true proofs of location;
  \item $FP$ [\%], the percentage of false proofs of location that are stored into the blockchain, with respect to the total number of false proofs of location;
  \item $ACC$ [\%], the percentage of true proofs of location stored into the blockchain and false proofs of location not stored into the blockchain, with respect to the total number of proofs of location.
\end{enumerate}
Simulation results for each $(P_1,P_2)$ combination have been obtained by averaging over 25 simulation runs. 

\begin{figure}[ht!]
  \centering
  \includegraphics[width=7cm]{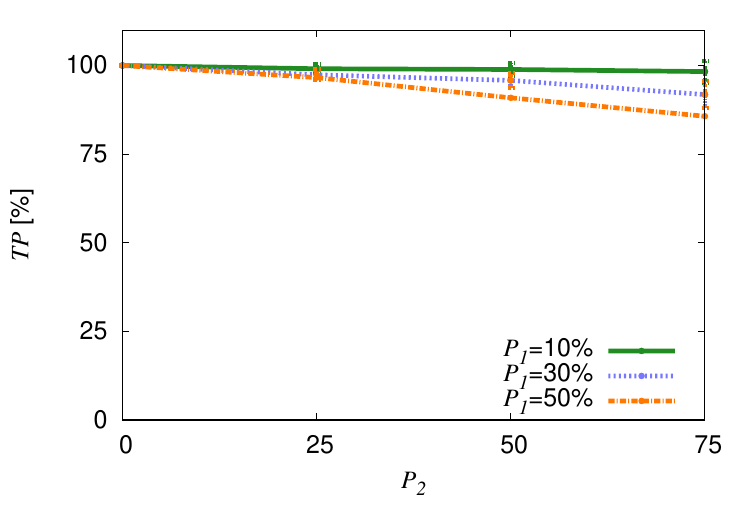}
  \includegraphics[width=7cm]{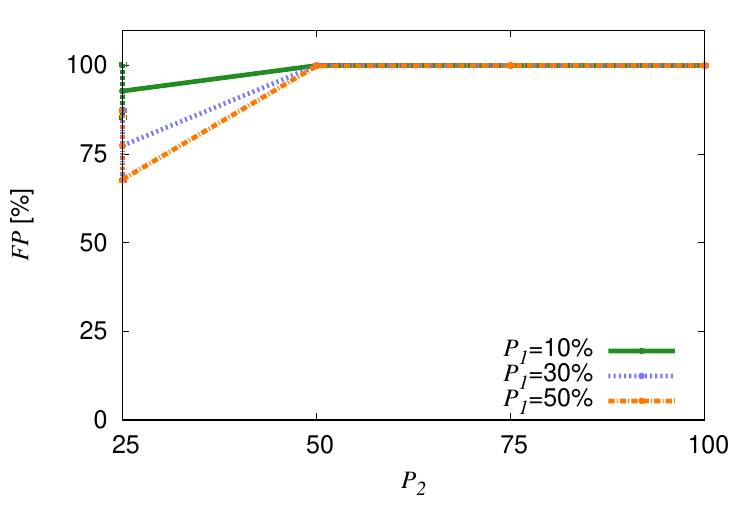}
  \includegraphics[width=7cm]{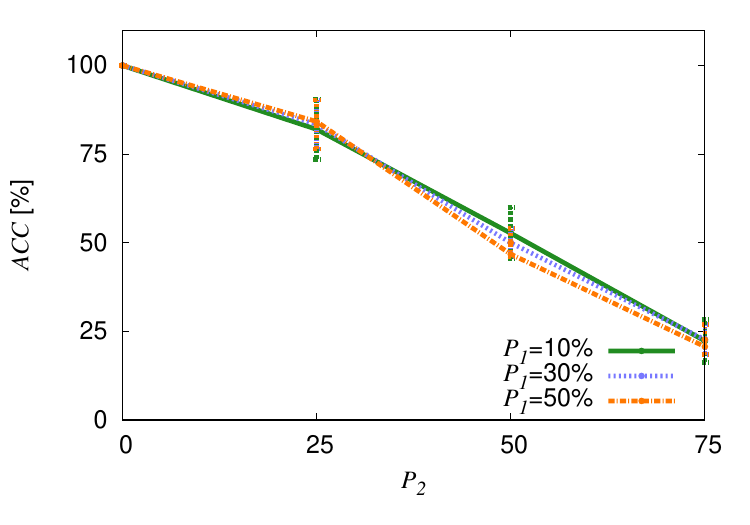}
  \caption{$TP$, $FP$ and $ACC$ for different values of $P_1$ and $P_2$, considering the full-cheating scenario.}
  \label{fig:fullCheatingResults}
\end{figure}

\begin{figure}[ht!]
  \centering
  \includegraphics[width=7cm]{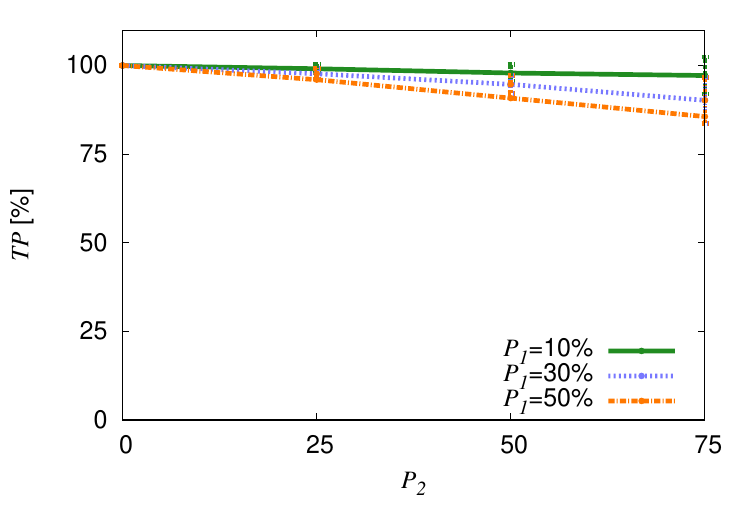}
  \includegraphics[width=7cm]{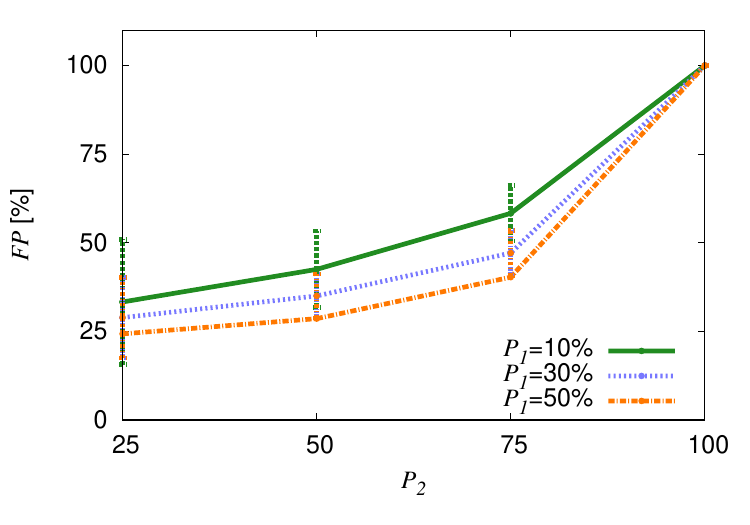}
  \includegraphics[width=7cm]{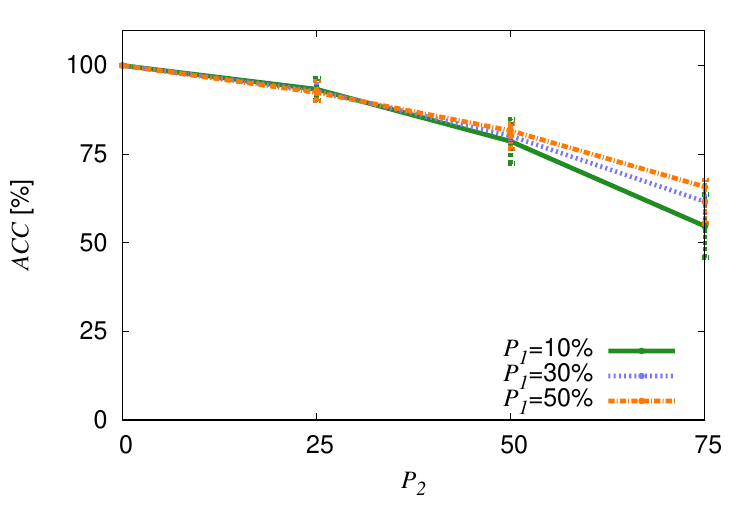}
  \caption{$TP$, $FP$ and $ACC$ for different values of $P_1$ and $P_2$, considering the less-cheating scenario.}
  \label{fig:lessCheatingResults}
\end{figure}

In Figures \ref{fig:fullCheatingResults} and \ref{fig:lessCheatingResults} the effects of full-cheating and less-cheating behaviors are illustrated. We can observe that with full cheating, even a reduced percentage of cheaters is able to sculpt false proofs of location into the blockchain, if not detected in time. 
Anyway, full cheating peers are easy to detect and isolate before they can cause a serious damage, because of their systematic discard of denial of location messages. 
The less cheating strategy, which is more difficult to detect, on the other hand requires the majority of Witnesses to be cheaters, in order to sculpt false proofs of location into the blockchain. 

We expect to achieve even better results with the adoption of the cheating detection strategy based on betweenness evaluation.

%%%%%%%%%%%%%%%%%%%%%%%%%%%%%%%%%%%%%%%%%%%
\section{Conclusion}
\label{sec:conclusion}

In this paper, we have presented a novel approach for producing proofs of location, \textit{i.e.}, digital certificates that attest someone's presence at a certain geographic location, at some point in time whereby LBSs can validate user locations.
We have illustrated a completely decentralized, blockchain-based peer-to-peer scheme that guarantees location trustworthiness and preserves user privacy, at the same time. We have analyzed the robustness of the proposed scheme against all major LBS-related attacks.
Furthermore, we have presented a preliminary simulation-based performance evaluation of the proposed scheme.

Regarding future work, we will study possible variants of the proposed scheme, for example the possibility that the proof-of-location service has a cost for the users and that block creators are rewarded.
Moreover, we plan to implement the proposed scheme in a plug-in for ADGT.js \cite{Brambilla2017}, our cross-platform, WebRTC-based realization of the ADGT peer-to-peer overlay protocol. To this purpose, we will adopt the Web Bluetooth API (https://webbluetoothcg.github.io/web-bluetooth/), a specification that allows web pages to discover and communicate with devices over the Bluetooth 4 wireless standard.

%%%%%%%%%%%%%%%%%%%%%%%%%%%%%%%%%%%%%%%%%%%
\section*{Acknowledgements}
The work of Michele Amoretti was partially supported by the University of Parma Research Fund - FIL 2016 - Project ``NEXTALGO: Efficient Algorithms for Next-Generation Distributed Systems'' and by the Future Technology Lab - Project ``MdM 4.0''.


%\bibliographystyle{IEEEtran}
\bibliographystyle{plainurl}
\bibliography{paperIWBS2018}

\end{document}